\documentclass[twocolumn,10pt,5p]{elsarticle}

\usepackage{atlasphysics} 
                           
\usepackage{subfigure}
\usepackage{mathrsfs}
\usepackage{graphicx}
\usepackage{multirow}
\usepackage{rotating}
\usepackage{xspace}
\usepackage{lineno}

\newcommand{\RP}{\ensuremath{R}-parity}

\def\lumi{4.4~\ifb}

\def\taumc{\ensuremath{\tau_{\rm MC}}}

\def\rdv{\ensuremath{r_{\rm DV}}}
\def\zdv{\ensuremath{z_{\rm DV}}}
\def\mdv{\ensuremath{m_{\rm DV}}}
\def\ntdv{\ensuremath{N^{\rm trk}_{\rm DV}}}

\def\zpv{\ensuremath{z_{\rm PV}}}

\def\mm{\rm mm}

\def\beq{\begin{equation}}
\def\eeq{\end{equation}}
\def\beqa{\begin{eqnarray}}
\def\eeqa{\end{eqnarray}}

\def\papertitle{Search for long-lived, heavy particles in final states
  with a muon and multi-track displaced vertex in proton-proton
  collisions at $\sqrt{s}=7$~$\mathrm{TeV}$ with the
ATLAS detector}

\usepackage{preprintcover}  

\PreprintCoverPaperTitle{Search for long-lived, heavy particles in final states
  with a muon and multi-track displaced vertex in proton-proton
  collisions at $\sqrt{s}=7$~$\mathrm{TeV}$ with the
ATLAS detector}

\PreprintIdNumber{CERN-PH-EP-2012-259}  
\PreprintCoverAbstract{
Many extensions of the Standard Model posit the existence of heavy
particles with long lifetimes. In this Letter,  results are presented
of a search for events containing one or more such particles, which decay
at a significant distance from their production point, using a final state containing charged hadrons and an associated muon. This analysis uses a data sample of proton-proton
collisions at $\sqrt{s}= 7$~$\mathrm{TeV}$ corresponding to an integrated luminosity of $\lumi$ collected in 2011 by the ATLAS detector operating at the Large Hadron Collider.
Results are interpreted in the context of \RP\ violating
supersymmetric scenarios. No events in the signal region are observed
and limits are set on the production cross section for pair production of
supersymmetric particles, multiplied by the square of the branching
fraction for a neutralino to decay to charged hadrons and a muon,
based on the scenario where both of the produced supersymmetric
particles give rise to neutralinos that decay in this way.  However,
since the search strategy is based on triggering on and reconstructing
the decay products of individual long-lived particles, irrespective of
the rest of the event, these limits can easily be reinterpreted in
scenarios with different numbers of long-lived particles per event.
The limits are presented as a function of neutralino lifetime, and for
a range of squark and neutralino masses.
}\PreprintJournalName{Physics Letters B}  

\begin{document}
\begin{frontmatter}

\title{\papertitle}
\author{The ATLAS Collaboration}

\begin{abstract}

Many extensions of the Standard Model posit the existence of heavy
particles with long lifetimes. In this Letter,  results are presented
of a search for events containing one or more such particles, which decay
at a significant distance from their production point, using a final state containing charged hadrons and an associated muon. This analysis uses a data sample of proton-proton
collisions at $\sqrt{s}= 7$~$\mathrm{TeV}$ corresponding to an integrated luminosity of $\lumi$ collected in 2011 by the ATLAS detector operating at the Large Hadron Collider.
Results are interpreted in the context of \RP\ violating
supersymmetric scenarios. No events in the signal region are observed
and limits are set on the production cross section for pair production of
supersymmetric particles, multiplied by the square of the branching
fraction for a neutralino to decay to charged hadrons and a muon,
based on the scenario where both of the produced supersymmetric
particles give rise to neutralinos that decay in this way.  However,
since the search strategy is based on triggering on and reconstructing
the decay products of individual long-lived particles, irrespective of
the rest of the event, these limits can easily be reinterpreted in
scenarios with different numbers of long-lived particles per event.
The limits are presented as a function of neutralino lifetime, and for
a range of squark and neutralino masses.

\end{abstract}

\end{frontmatter}

\section{Introduction}
\label{sec:Introduction}
Several extensions to the Standard Model predict the production of heavy particles with lifetimes that may
be of order picoseconds up to about a nanosecond~\cite{Fairbairn:2006gg}, at the Large Hadron Collider (LHC).
One such scenario is gravity-mediated
supersymmetry (SUGRA~\cite{sugra}) with  \RP\ violation (RPV)~\cite{Barbier:2004ez,Allanach:2006st}. If the results of
$R$-parity conserving supersymmetry
searches at the LHC (see, for example,
Refs.~\cite{:2012pq,Chatrchyan:2012te}) continue to disfavour light superpartners, then scenarios in which $R$-parity is violated may be required if
supersymmetry is the solution to the hierarchy problem~\cite{Graham:2012th}. The present (largely indirect) constraints on RPV
 couplings~\cite{Barbier:2004ez,Allanach:2006st} would allow the decay of the lightest supersymmetric particle as it traverses
a particle detector at the LHC. Signatures of heavy, long-lived particles also feature in models of gauge-mediated supersymmetry
breaking~\cite{Dimopoulos:1996vz}, split-supersymmetry~\cite{Hewett:2004nw}, hidden-valley~\cite{Strassler:2006im},
dark-sector gauge bosons~\cite{Schuster:2009au} and stealth supersymmetry~\cite{Fan:2011yu} though some of these models are disfavoured~\cite{Arbey:2011ab} by the
recent results on the search for the Higgs boson~\cite{:2012gk,:2012gu}.

\begin{figure}[!hbtp] \centering \includegraphics[width=5cm]{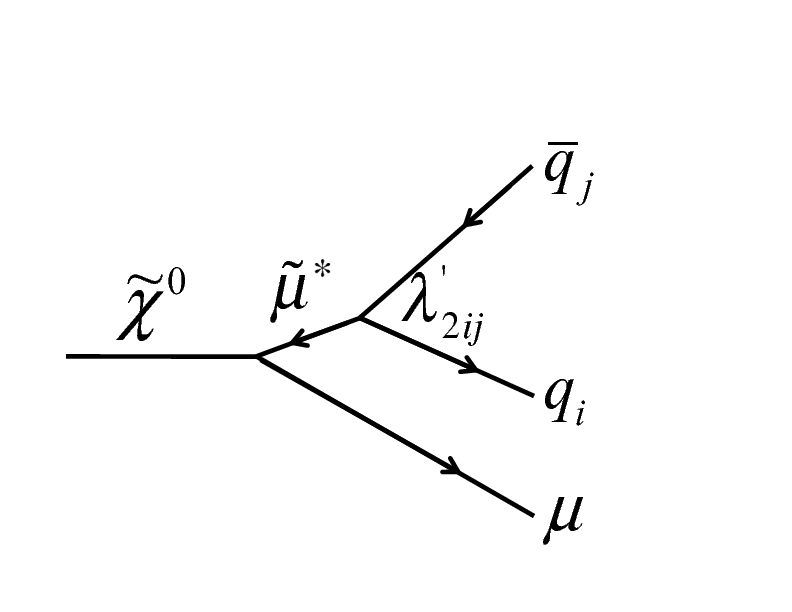}
\caption{\label{fig:RPV-diagram}
Diagram of the massive lightest neutralino $\tilde\chi^0$ decaying into a muon and two jets via a virtual smuon, with lepton-number and
$R$-parity violating coupling $\lambda'_{2ij}$.}
\end{figure}

This Letter presents the results of a search for the decay of a heavy particle, producing a multi-track vertex that contains a muon with high transverse momentum (\pt) at a distance between millimetres and tens of centimetres from the $pp$ interaction point.  As in the earlier ATLAS work~\cite{Aad:2011zb}, the results are interpreted in the context of an $R$-parity violating supersymmetric scenario. In this model, the signature under consideration corresponds to the decay of the lightest supersymmetric particle, resulting in a muon and many high-\pt~charged tracks originating from a single displaced vertex (DV). This can arise from a diagram such as that shown in Fig.~\ref{fig:RPV-diagram}, where the decay occurs due to the  non-zero RPV coupling $\lambda'_{2ij}$.

The CMS collaboration has performed a search for a related signature,
where a long-lived neutral particle decays to a final state containing
two leptons, using $5 \mathrm{pb}^{-1}$ of $p p$ collisions at
$\sqrt{s}=7~\tev$~\cite{CMS_LLP}\footnote{Following submission of our paper, this paper was submitted addressing the
same physics, but for a different final state.}.
Similar searches have also been performed by the {D\O} collaboration at the Tevatron
with $\sqrt{s}=1.96~\tev$ $p\bar p$ collisions~\cite{Abazov:2006as},
who also study decays to a $b\bar b$ pair~\cite{D0:2009},   
and at LEP~\cite{Heister:2002vh}.
The result reported here improves on the previous ATLAS search for this signature~\cite{Aad:2011zb} in several ways. The dataset is around $100$ times larger, corresponding to an integrated luminosity of
$\lumi$~\cite{lumid}. Furthermore, the signal efficiency is increased
by improving the reconstruction efficiency 
for reconstructing tracks that do not originate from the primary vertex. The
dependence on simulation to estimate background levels is also
removed.  Finally, while the previous analysis did not require muons to
be associated vertices, this association requirement is now imposed,
which greatly simplifies both the analysis itself, and also the re-interpretation of these
results for scenarios containing different numbers of long-lived particles.

\section{The ATLAS detector}
\label{sec:detector}

The ATLAS detector~\cite{ATLAS:detector2008} is a multipurpose apparatus at the
LHC. The detector consists of several layers of subdetectors. From the interaction
point (IP) outwards there is an inner detector (ID), measuring the tracks of charged particles, electromagnetic and hadronic
calorimeters, and an outer muon spectrometer (MS).

The ID is immersed in a 2~T axial magnetic field, and extends from a radius\footnote{ATLAS uses a right-handed coordinate system with its origin at the nominal
IP in the centre of the detector and the $z$-axis along the beam pipe. The $x$-axis points from the IP to the centre of the LHC ring, and the $y$-axis points upward. Cylindrical coordinates $(r,\phi)$ are used in the transverse plane, $\phi$ being the azimuthal angle around the beam pipe. The pseudorapidity is defined in terms of the polar angle $\theta$ as $\eta=-\ln\tan(\theta/2)$.} of about 45~mm to 1100~mm and to $|z|$ of about 3100~mm. It provides tracking and vertex information for charged particles within the pseudorapidity region $|\eta| < 2.5$. At small radii, silicon pixel layers and stereo pairs of silicon microstrip detectors provide high resolution pattern recognition. The pixel system consists of three barrel layers, and three forward disks on either side of the interaction point (the beam pipe envelope is at a radius of about 30 mm). The barrel pixel layers, which are positioned at radii of 50.5~mm, 88.5~mm, and 122.5~mm are of particular relevance to this work. The silicon microstrip tracker (SCT) comprises four double layers in the barrel, and nine forward disks on either side. A further tracking system, a transition-radiation tracker (TRT), is positioned at larger radii. This device is made of straw-tube elements interleaved with transition-radiation material; it provides coverage within the region $|\eta| < 2.0$.

The calorimeter provides coverage over the pseudorapidity range $|\eta| < 4.9$. It consists of a lead/liquid-argon electromagnetic calorimeter, a hadronic
calorimeter comprising a  steel and scintillator-tile system in the barrel region and a
liquid-argon system with copper and tungsten absorbers in the end-caps.

Muon identification and momentum measurement is provided by the MS. This device has a
coverage in pseudorapidity of $|\eta|<2.7$ and is a three-layer system of gas-filled
precision-tracking chambers. The pseudorapidity region $|\eta|<2.4$ is additionally covered by
separate trigger chambers, used by the hardware trigger for the first level of triggering (level-1).
The MS is immersed in a magnetic field which is produced by a set of toroid magnets, one for the barrel and one each for the two
end-caps. Online event selection is performed with a three-level trigger system. It comprises a
hardware-based level-1 trigger, which uses information from
the MS trigger chambers and the calorimeters, followed by two software-based trigger levels.

\section{Data and simulation}
\label{sec:data}

The data used in this analysis were collected between March and October, 2011. After the application of beam, detector, and data-quality
requirements, the integrated luminosity considered corresponds to $\lumi$, with an uncertainty of $\pm 3.9$\%~\cite{lumid}.

Signal Monte Carlo (MC) events are generated with {\sc Pythia}~6~\cite{pythia}, using the
MRST LO** ~\cite{Sherstnev:2009} set of parton distribution functions (PDFs).  Processes are simulated
in which a $\tilde{q}\tilde{q}$,
$\overline{\tilde{q}}\,\overline{\tilde{q}}$, or $\tilde{q}\overline{\tilde{q}}$ pair is produced in the $pp$
collision, with each squark (antisquark) decaying into a long-lived lightest neutralino and a quark (antiquark).
Degeneracy of the first and second generations and for the left-handed and right-handed squarks
is assumed. The masses of the gluino, sleptons and third-generation squarks are set at such a high value (5~TeV) that they are not directly produced
in the supersymmetry scenario considered here.

The parameter settings for the samples of signal MC events are summarised in
Table~\ref{tab:sigMC}. The chosen values
of squark and neutralino masses correspond to a wide range in the quantities to which the signal
efficiency is most sensitive: neutralino speed and final-state multiplicities (see
Section~\ref{sec:eff}). The Higgsino-gaugino mixing determines the neutralino mass values. The signal MC
samples are labelled in Table~\ref{tab:sigMC} according
to the squark mass and
neutralino mass, respectively: MH (medium-mass squark, heavy
neutralino), 
ML (medium-mass squark, light neutralino), and HH (heavy
squark and heavy neutralino). It is assumed that all RPV couplings other than $\lambda'_{211}$ are zero. This
leads to each neutralino decaying to $\mu^- u
\bar d$ (or the charge-conjugate state). Decays of neutralinos to a neutrino and jets final state, which would also occur via a non-zero
$\lambda'_{211}$, are not considered in this work.

The {\sc PROSPINO}~\cite{Beenakker:1996ed} program is used to calculate the signal cross section (hereafter referred to as the supersymmetry production cross section) at
next-to-leading
order; cross section  values are listed in Table~\ref{tab:sigMC}. The
CTEQ6.6 PDF set~\cite{Nadolsky:2008zw} is used. Owing to differences
in the  predicted  cross sections for the $\tilde q \tilde q$ and $\tilde{q} \bar{\tilde{q}}$
processes from {\sc Pythia} and  {\sc PROSPINO},  and the possible sensitivity of the neutralino-speed distribution to
the assumed process,  signal samples are generated
for each process separately with {\sc Pythia} and reweighted according to the relative cross sections of the two
processes as estimated with {\sc PROSPINO}. The reweighting procedure, however, has a minimal effect on the distributions of physical quantities
used in this work and typically leads to changes of much less than 1\%.

\begin{table}
\begin{center}
\renewcommand\arraystretch{1.1}
\begin{tabular}{|c|c|c|c|c|c|c|}
\hline
Sample & $m_{\tilde{q}}$  & $\sigma$
       & $m_{\tilde{\chi}^0_1}$
       &  $\left<\gamma\beta \right>_{\tilde{\chi}^0_1}$
       & $c\taumc$
       & $\lambda_{211}^{\prime}$
        \cr
    &  [GeV]           & [fb]     & [GeV]  &  & [\mm] & $\times 10^{-5}$ \cr
\hline
MH & 700  & 66.4               & 494   & 1.0  &  78 & $0.3$ \\
ML & 700  & 66.4               & 108   & 3.1  &  101 & $1.5$\\
HH & 1500 & 0.2                & 494   & 1.9  &  82 & $1.5$\\
\hline
\end{tabular}
\caption{\label{tab:sigMC}
The parameter values for the three signal MC samples used in this work: the assumed squark mass, production cross section
(calculated with
{\sc PROSPINO}), neutralino mass, average value of the Lorentz boost factor (from {\sc Pythia}),
average proper decay length, and value of $\lambda_{211}^{\prime}$
coupling used in the generation of the signal MC samples.}
\end{center}
\end{table}

Each generated event in the signal samples is processed with the {\sc Geant}4-based~\cite{geant4} ATLAS detector
simulation~\cite{ATLASgeant4} and treated in the same way as the
collision data.  The samples include a realistic modelling of the effects of multiple $pp$ collision per bunch crossing
(pile-up) observed in the data, obtained by overlaying simulated
Minimum Bias events generated 
using {\sc Pythia}, on top of the hard
scattering events, and reweighting events such that the distribution
of the number of interactions per bunch crossing matches that in the data..

\section{Vertex reconstruction and event selection}
\label{sec:cuts}

It is required that the trigger identifies a muon candidate with transverse momentum \pt~$> 40$~\gev\
and $|\eta|<1.05$, i.e., in the barrel region of the MS. The latter
selection is necessary since there is no ID track requirement in the
trigger, and the trigger rate arising from muons with badly measured
momenta in the (larger $|\eta|$) end-cap region, where the magnetic
field is highly complex, would have consumed too much bandwidth to be
viable in the 2011 running conditions.

Owing to the higher instantaneous luminosity of the LHC in 2011 compared to that in 2010, most events contain more than one primary vertex (PV). The PV with the highest sum of $p_{\mathrm{T}}^2$ of the tracks associated to it is required to have at least five tracks and a $z$ position in the range $|\zpv|<200~\mm$. In order to reduce cosmic-ray background, events are rejected if they contain two muons which appear back-to-back. A selection $\sqrt{(\pi - (\phi_1 - \phi_2))^2 - (\eta_1 + \eta_2)^2}>0.1$ is applied where $\phi_1$, $\phi_2$, $\eta_1$ and $\eta_2$ are the azimuthal angles and pseudorapidities of the two reconstructed muons.

A muon candidate is required to have been reconstructed in both the MS and the ID with transverse momentum \pt~$> 50$~GeV (which is well into the region
where the trigger efficiency is approximately independent of the muon
momentum) and $|\eta|<1.07$. The impact parameter relative to the transverse position of the
primary vertex ($d_0$) is required to satisfy $|d_0|> 1.5$~mm. To
ensure that
the muon candidate is associated with the muon that satisfied the trigger requirement, the selection $\sqrt{(\Delta\phi)^2 + (\Delta\eta)^2}
< 0.15$ is imposed, where $\Delta\phi$ ($\Delta\eta$) is the difference between the azimuthal angle (pseudorapidity) of the reconstructed muon
and that of the muon identified by the trigger. The ID track associated with the muon candidate is required to have at least six
SCT hits, and at most one SCT hit that is expected but not found i.e. in an active detector element with no known read-out problems. Furthermore, the track must
satisfy an $|\eta|$-dependent requirement on the number of TRT hits. No pixel-hit requirements are applied to the muon track. The combination of requirements described above
is referred to as the muon-selection criteria.

\begin{figure}[!hbtp]
\centering
\includegraphics[width=9cm]{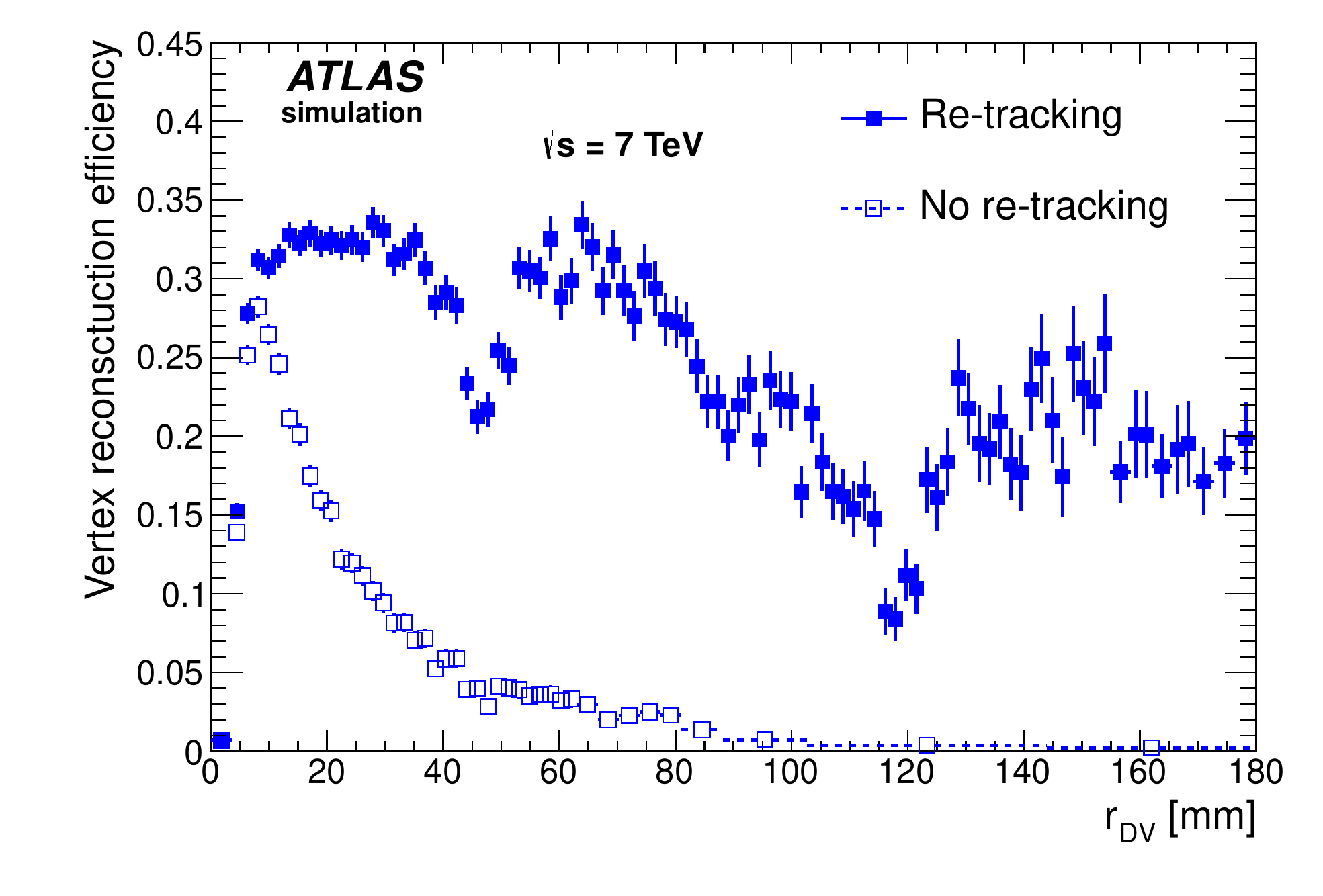} \\
\caption{\label{fig:retrack_improve}
The vertex reconstruction efficiency as a function of $r_{\mathrm{DV}}$ for the MH sample, with and without re-tracking. All selection criteria, except for the material veto, have been applied.}
\end{figure}

In order to reconstruct a DV it is first necessary to select high quality tracks.
This is done by requiring that candidate tracks have two or more associated SCT hits
and a value of $|d_0|$ greater than 2~mm. Studies made with the MC simulation show that 98\% of all tracks originating from the primary $pp$
interaction are rejected by the selection on $|d_0|$. Standard ATLAS tracking~\cite{ATLAS:detector2008} is based on three passes in which the initial track candidates
are formed in different ways:
initially
found in the silicon detectors, initially found in the TRT detector, and only found in the TRT detector. The algorithms all assume that
tracks originate from close to the PV, and hence have reduced reconstruction efficiency for signal tracks which originate at a DV. To
counter this problem and recover some of these lost tracks, the silicon-seeded tracking algorithm is re-run with looser requirements on the
radial and $z$ impact parameters, and on the number of detector hits that can be shared among more than one seed track. To reduce the rate of false
seed tracks, it is required that these additional tracks have \pt~$>1$~GeV, which is greater than the standard-tracking requirement of
\pt~$>400$~MeV.  This procedure is termed ``re-tracking''. Fig.~\ref{fig:retrack_improve} shows the efficiency for vertex reconstruction (described later) as a function of
the radial position of the vertex $r_{\mathrm{DV}}$ when using standard tracking and re-tracking for sample MH. As can be seen, there is a substantial improvement due to re-tracking at
values of $r_{\mathrm{DV}}$ greater than about 10~mm. The dips in the plot correspond to losses in efficiency for decays immediately before a pixel layer, where many tracks from the
vertex have shared pixel hits and therefore fail the selection.

Using an algorithm based on the incompatibility-graph approach~\cite{Das:1973}, DVs are sought with the
selected tracks. The method adopted is similar
to that used in Ref.~\cite{vtxNote}. The algorithm starts by finding two-track seed vertices from all
pairs of tracks;
those that have a vertex-fit $\chi^2$ of less than 5.0 per degree of freedom are retained. A seed vertex is rejected if at least one of its
tracks has hits between the vertex and the
PV. Multi-track vertices are formed from combinations of seed vertices in an iterative process. The
following method is used to do this. If a track is assigned to two different vertices, the action taken
depends on the distance $D$ between the vertices. If $D < 3 \sigma_D$, where
$\sigma_D$ is the estimated
uncertainty on $D$, a single vertex is formed from the tracks of the two vertices. Otherwise, a track is
associated with that vertex for which the track has the smaller $\chi^2$ value. In the event of the
$\chi^2$ of a track relative to the resulting vertex exceeding 3.0 per degree of freedom, the track is removed from the vertex, and the vertex is refitted. This process continues until no track is associated with more than one vertex. Finally, vertices are combined and refitted if
they are separated by less than 1~mm.

To ensure a good quality DV fit, the $\chi^2$ per degree of freedom of the fit must be less
than 5. Furthermore, the DV position is required to be in the fiducial region, defined as $|\zdv|<300$~mm and
$\rdv<180$~mm, where $r_{\mathrm{DV}}$ and $z_{\mathrm{DV}}$ are the radial and
longitudinal vertex positions with respect to the origin. To minimise background coming from the PVs, the
transverse distance
$\sqrt{(x_{\rm DV} - x_{\rm PV})^2 + (y_{\rm DV} - y_{\rm PV})^2}$ between any of the PVs and
the DV must be at least 4~mm. Here $x$ and $y$ are the transverse coordinates of a given vertex, with the subscripts PV and DV
denoting the type of vertex. The number of tracks \ntdv\ associated with the DV is required to be at
least five. This reduces background from random
combinations of tracks and from material interactions. Background due to
particle interactions with material
is suppressed by requiring $\mdv > 10$~\gev. Here, \mdv\ is the invariant mass of the set of tracks
associated with the DV, using the charged-pion mass hypothesis for
each track. Candidate vertices
which pass (fail) the $\mdv > 10$~\gev\ requirement are hereafter referred to as being high-\mdv\ (low-\mdv) vertices.

The typical position resolution of the DV in the signal MC samples is tens of microns for \rdv\ and about 200~$\mu$m for \zdv\ near
the interaction point. For vertices beyond the outermost pixel layer, which is located at a radius of 122.5~\mm, the typical
resolution is several hundred microns for both coordinates.

Low-\mdv\ (\mdv$<4\mathrm{GeV}$) vertices from particle--material interactions are abundant in regions in which
the detector material is dense. High-\mdv\
background may arise from the random spatial coincidence of a material-interaction
vertex with a high-\pt~track, especially when this track and the
particle that initiated the material-interaction vertex originate from different primary interactions,
resulting in a large angle between their momentum vectors.

To reduce this source of background, vertices that are reconstructed in regions of
high-density material are removed. The high-density material was mapped using
low-\mdv\ material-interaction candidate vertices in data and true material-interaction vertices
in minimum-bias MC events. The
\zdv\ and \rdv\ positions of these vertices are used to make a two-dimensional material-density distribution with a
bin size of 4~mm in \zdv\ and 1~mm in \rdv. It has been demonstrated~\cite{vtxNote} that the detector
simulation describes the positions of pixel layers and associated material reasonably well, while the simulated
position of the beampipe is shifted with respect to the actual position. The use of data events to
construct the material map therefore ensures that the beampipe material is correctly mapped, while the use of the simulation
provides a high granularity of the map at the
outer pixel layers, where material-interaction vertices in the data are
comparatively rare. Material-map bins with vertex density greater than an \rdv- and \zdv-dependent density
criterion are characterised as high-density-material
regions. These make up 34\% of the volume  at $|\zdv|<300$~mm,
$\rdv<180$~mm.

To ensure that the muon candidate is associated with the reconstructed DV,
the distance of closest approach of the muon with respect to the DV is
required to be less than $0.5$~mm.  This requirement ensures that the
vertex that we reconstruct gave rise to the muon that triggered the
event, and so the selection efficiency for each neutralino decay is independent
of the rest of the event.  This facilitates a straightforward
calculation of the event selection efficiency for scenarios with
different numbers of long-lived neutralinos in the event.
The aforementioned
selections are collectively referred to as the vertex-selection
criteria.  Events containing one or more vertices passing these criteria
are accepted.

\section{Signal efficiency}
\label{sec:eff}

The signal efficiency depends strongly on the efficiencies for track reconstruction and track selection, which are affected by several
factors: (1) The number of tracks originating from the DV and their total invariant mass increase with the neutralino mass. (2) More tracks
fail the $|d_0|>2$~mm requirement for small \rdv\ or if a neutralino with large \pt~has a small decay length, thereby leaving its daughters pointing
back closer to the PV. (3) The efficiency for reconstructing tracks decreases with increasing values of $|d_0|$. Because the MH and
HH samples have the same neutralino mass, but different boosts, a cross--check of the efficiency estimation procedure can be made.  The HH
sample is reweighted vertex-by-vertex such that the $\beta\gamma$ versus $\eta$ distributions match the MH sample. It was checked that they have the same efficiency
as a function of decay position.

\begin{figure}[!hbtp]
\centering
\includegraphics[width=9cm]{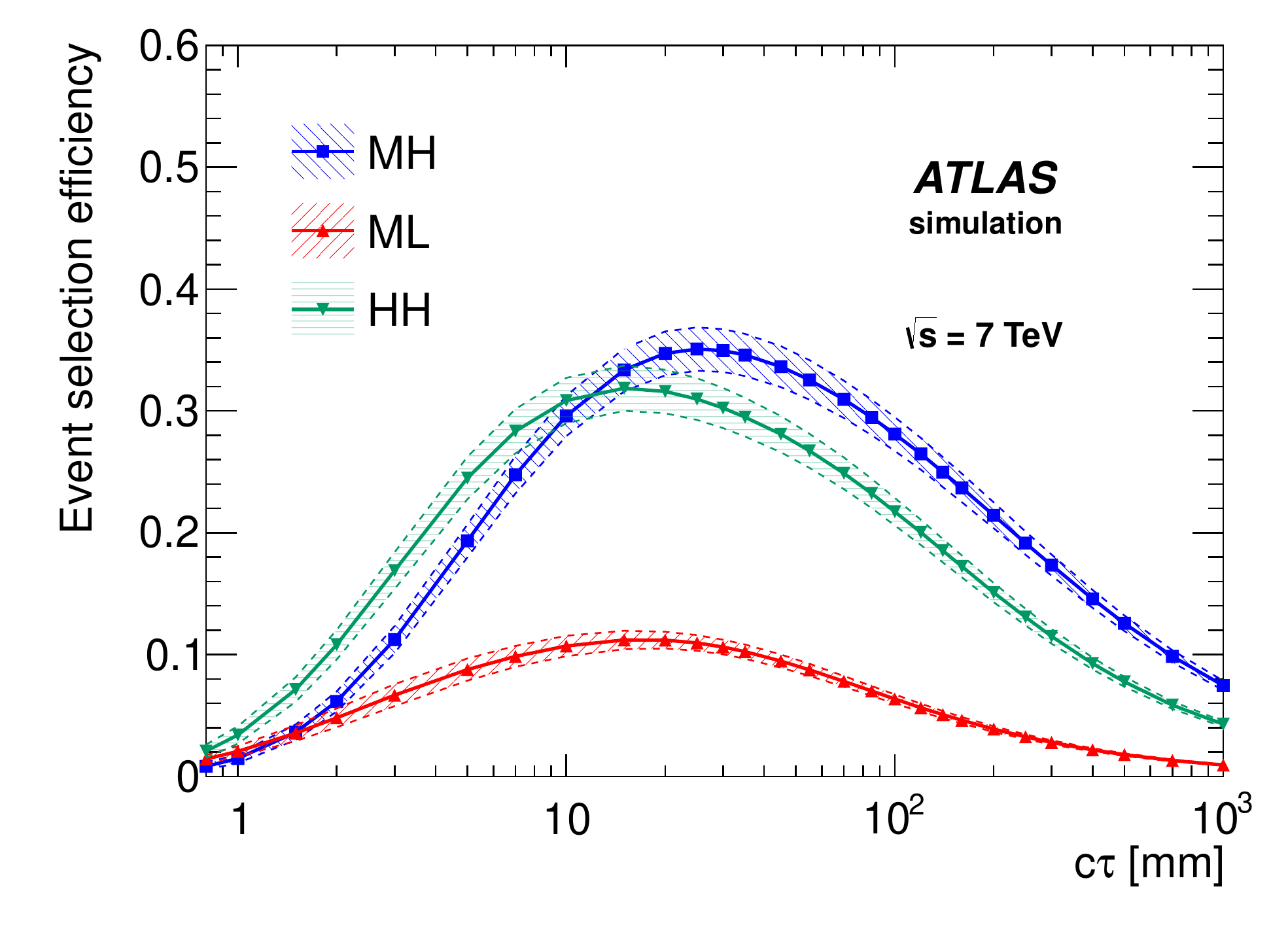} \\
\caption{\label{fig:eff}
The event selection efficiency as a function of $c\tau$ for the three
signal samples. The total uncertainties on the efficiency are shown as bands; the estimation of the uncertainties
is discussed in Section~\ref{sec:syst}.}
\end{figure}

\begin{figure}[!hbtp]
   \centering
   \includegraphics[width=9cm]{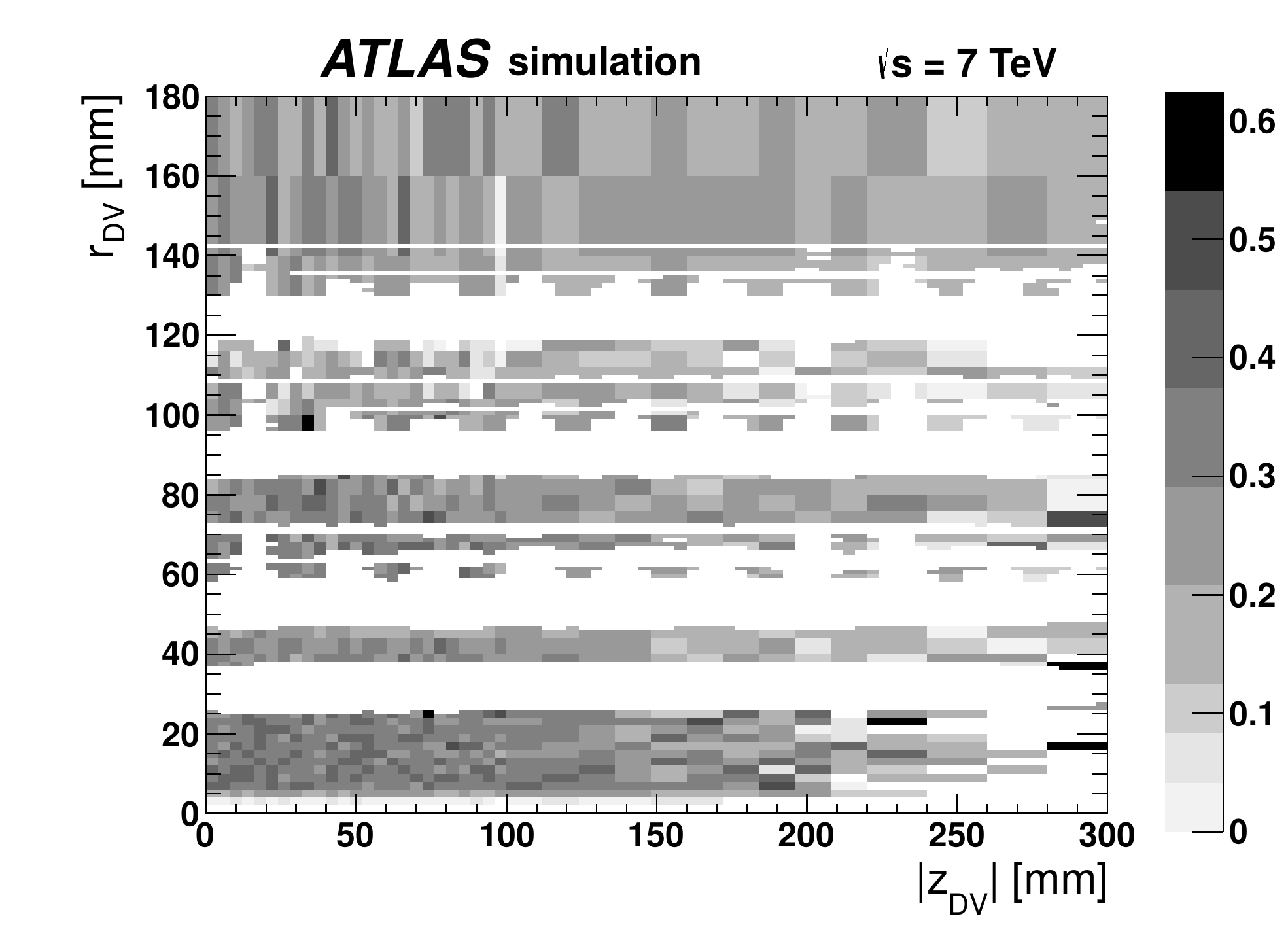}
   \caption{\label{fig:eff-2d}
	The efficiency as a function of \rdv\ and \zdv\ for vertices in
the signal MC sample~MH. The blank areas represent regions of dense
	material which are not considered when looking for
	DVs. Bin-to-bin fluctuations due to limited MC sample size are
	compensated for by recalculating this map many times, varying
	each bin content within its statistical uncertainty.}
\end{figure}

The total efficiency for each of the signal MC samples is shown in Fig.~\ref{fig:eff} as a function of $c\tau$. This is an overall event efficiency, derived from the determination of (the single) vertex efficiency, and is based on having two displaced vertices per event in the signal MC. This efficiency as a function of $c\tau$ is calculated from the samples generated with a single-$c\tau$ value (see Table~\ref{tab:sigMC}) using a two-dimensional map in $(|\zdv|, \rdv)$ (Fig.~\ref{fig:eff-2d}) and distributions of decay positions for different values of $c\tau$. Events are reweighted according to their probability of being found at different positions in the map, where the probability depends on $c\tau$. Figure~\ref{fig:eff} includes systematic corrections and uncertainties that are discussed in Section~\ref{sec:syst}.

\section{Background estimation}
\label{sec:bgd}

Spurious high-$m_{\mathrm{DV}}$, high-multiplicity vertices in non-material regions could come from one of two sources: (1) Purely random combinations of tracks
(real or fake). This type of background is expected to form the largest contribution at small radii where the track density is highest. (2) Vertices from real
hadronic interactions with gas molecules. Although most of these events will have masses below the $10~\gev$ requirement, the high-mass
tail of the distribution indicates a potential background, in particular if the vertex is crossed by a random track (real or fake) at large
crossing angle.

The first source of background mentioned above gives rise to displaced vertices inside the beampipe (where there are few real hadronic interactions due to the
good vacuum, but where random combinations are
expected to dominate). 
The expected contribution from this source of background is evaluated
using a large sample of jet-triggered events, and examining the ratio
of the number of vertices with $m_{DV} > 10 \mathrm{GeV}$ to the
number in a mass control region, $4\mathrm{GeV} < \mdv <
10\mathrm{GeV}$.  Since the number of random combinations of tracks
depends primarily on the pile-up conditions, and is therefore
independent of the type of event that fired the trigger, this ratio
can also be applied to the number of vertices in the mass control
region in muon-triggered events.  All of this information is
used as input to a simple maximum likelihood fit, which yields a
background expectation of $0.00^{+0.06}_{-0}$ vertices passing our
cuts, that arise from random combinations of tracks.

To estimate the second source of background i.e. vertices arising from real hadronic interactions, including those crossed by tracks from different PVs,
two $m_{\mathrm{DV}}$-distributions are constructed.
One distribution is formed from vertices which do not include a so-called large-angle track, i.e. the average angle between a constituent track and the
rest of the tracks
in the vertex must be less than one radian.
The four-momentum of a randomly-selected track in the event is added
to the vertex to build the second distribution. Both of
these distributions are taken from a large control sample of jet-triggered events. The relative
normalisations of these
distributions are determined using an estimate of the ``crossing probability''. This is obtained by fitting a function
representing a $K_S$ mass peak to the invariant mass distribution of
all two-track combinations within three-track vertices, and comparing
the integral of this function (which is a measure of how many real
$K_S$ were crossed by a random track to form a three-track vertex), to
the number of two-track $K_S$ mesons observed.

With the shape of the $m_{\mathrm{DV}}$-distribution determined in this manner, the absolute normalisation for the muon-triggered sample can be
found. It is extracted from a maximum likelihood fit, which takes as input the numbers of vertices in each material layer (i.e., the beam
pipe and three pixel detector layers\footnote{A detailed study of the material distribution within the ID
is provided in Ref.~\cite{vtxNote}.}), and the numbers of low-track-multiplicity
vertices in the regions in which there is no material, which are termed ``air gaps". It has been empirically observed that in
each material layer or air gap, the relation between the numbers of $n$-track,
$(n+1)$-track, $(n+2)$-track vertices etc., is well modelled by an
exponential function. Furthermore, the ratio of the number of
$n$-track vertices in a material layer to the number in the air gap
immediately outside it, is consistent between several different
control samples.  These relations can therefore be used in the fit
to predict the number of vertices with at least five tracks in the air
 gaps. This procedure was tested on various control samples, giving
 consistent results for the $\mdv < 10\mathrm{GeV}$ control region in
 all cases. For instance, in events that fail the
 muon trigger requirement, the fit predicted $105.4 \pm 5.7$ events,
 whereas 110 events were observed. 
The \mdv\ distribution, scaled by the prediction, is then integrated to obtain an
estimate of the number of vertices with $m_{\mathrm{DV}}>10$ $\mathrm{GeV}$,
and a final scale factor is applied, based on the fraction of events
in our muon-triggered sample that pass the offine muon requirements.  
Systematic uncertainties are evaluated by varying the relevant
input parameters (the crossing probability, the parameters of the
exponential function and material-to-airgap ratio, and the final scale
factor) within their respective uncertainties, and taking the largest deviation.
Finally, the background from hadronic interactions in the air gaps is
estimated to be $3.7^{+4.4}_{-3.7}\times 10^{-3}$ vertices.
Note that this is somewhat conservative, as the ``crossing
probability'' used here is derived from tracks crossing $K_S$
vertices, which have worse position resolution than vertices with four
or more tracks.

The total background estimate is therefore taken to be $4^{+60}_{-4}\times 10^{-3}$ vertices.

\section{Systematic corrections and uncertainties}
\label{sec:syst}

Several categories of uncertainties and corrections are
considered. The values of the uncertainties depend on the 
neutralino lifetime and the signal sample
under investigation. However, they can be ordered in approximately
decreasing size.  Uncertainties on the 
muon reconstruction efficiency are largest (typically $\approx 7\%$),
followed by trigger ($\approx 6\%$) and tracking ($\approx 4\%$)
efficiency uncertainties. The relative luminosity uncertainty is
$3.9\%$, while the uncertainty due to limited numbers of events in the
signal samples is around $1.5\%$.  Other systematic effects, such as
the uncertainty on the modelling of pile-up, contribute less than
$1\%$ to the total uncertainty.  The effect on the efficiency
associated with PDF uncertainties is negligible.

To estimate the uncertainty on the efficiency as a function of $c\tau$ (Fig.~\ref{fig:eff}) from finite MC sample size, the efficiency in each bin of the two-dimensional efficiency map (Fig.~\ref{fig:eff-2d}) was varied randomly within its statistical uncertainty and the total efficiency was repeatedly re-measured. The spread in those results is taken as the contribution to the systematic uncertainty on the efficiency at each $c\tau$ value.

To compensate for the different $z$-distributions of the primary vertices in data and MC simulation, a weight is applied to each
simulated event such that the reweighted $z_{\mathrm{PV}}$ distribution matches that in data. This weight is applied to both the numerator and
denominator in the efficiency calculation.

Similarly, a weight is applied to each simulated event such that the distribution of $\left< \mu \right>$, the average number of interactions per bunch-crossing, matches that in data. An uncertainty associated with this procedure is estimated by scaling the $\left< \mu \right>$ values used as input to this correction calculation by a factor 0.9 (motivated by taking the difference between scaling the MC simulation by $\left< \mu \right>$ calculated from luminosity measurements and by using the number of reconstructed PVs in data and MC simulation). The difference in the efficiency is evaluated as a function of $c\tau$, and is applied as a symmetric systematic uncertainty.

A trigger efficiency correction and its associated systematic uncertainty are derived from a study of $Z\rightarrow \mu^+\mu^-$ events in which
at least one of the muons was selected with a single-muon trigger. The trigger efficiency for selecting a signal event is
about 90\%.  The ratio of the trigger efficiencies in
data and simulation is applied as a correction factor. Furthermore, the statistical uncertainty on this ratio is added in quadrature to the
differences in trigger efficiency between the $Z\rightarrow \mu^+\mu^-$ sample and the signal samples to estimate a systematic uncertainty.

The modelling of track reconstruction efficiency for prompt tracks has
been extensively studied in ATLAS~\cite{TrackReco}.  In order to
estimate the systematic uncertainty on the track reconstruction for
secondary tracks, a study is performed where the number of $K_S$ vertices in data and simulation are compared over a range of radii and
$\eta$-values. Based on the outcome of this study, some fraction of tracks in the signal MC are randomly removed from the input to the vertexing algorithm, and
the same procedure is used to obtain efficiency-vs-$c\tau$ is performed. The difference between this and the nominal efficiency at each $c\tau$ point is
taken as the systematic uncertainty.

Since the distribution of the true values of $d_0$ for cosmic-ray muons is flat over the limited $d_0$ range considered here, as predicted by cosmic-ray simulation, the shapes of the  measured (and simulated) $d_0$-spectra of such muons are used to determine the accuracy to which the simulation reproduces the muon-finding efficiency as a function of $d_0$. The ratio of the number of observed cosmic-ray muons (recorded during 2011) and the simulated muon-finding efficiency is studied as a function of $d_0$. The degree to which the distribution is not flat is used to estimate the $d_0$-dependent muon-finding efficiency, which is used for the results in Figs.~\ref{fig:eff} and~\ref{fig:eff-2d}. The uncertainties on the efficiency (largely due to the limited statistical precision of the cosmic-ray background sample) are propagated to uncertainties of between $\pm 3.5$\% (MH) and $\pm 8$\% (ML) on the signal reconstruction efficiency, depending on the signal sample.

\section{Results}
\label{sec:results}

\begin{figure}[!hbtp]
\begin{center}
  \includegraphics[width=9cm]{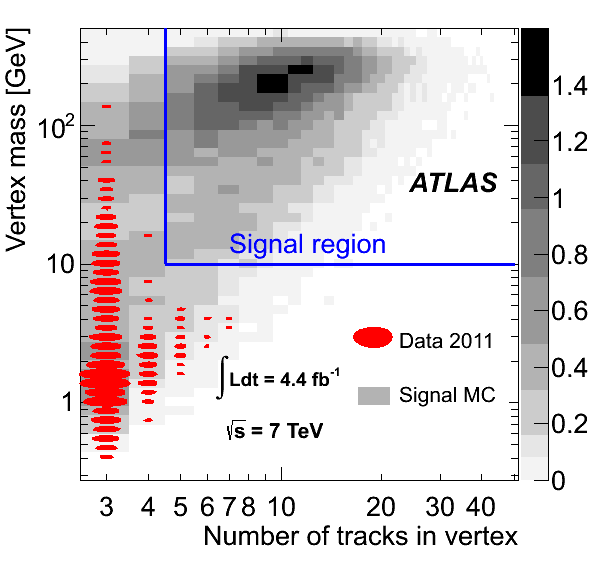}
  \caption{\label{fig:massVsMult}
Vertex mass (\mdv) vs. vertex track multiplicity (\ntdv) for DVs in events with no muon requirements other than the trigger, and all other selection criteria except the \mdv\ and \ntdv\ requirements. Shaded bins show the distribution for the signal MC
MH sample (see Table~\ref{tab:sigMC}), and data are shown as filled ellipses, with the area of the ellipse proportional to the number of
vertices in the corresponding bin.}
\end{center}
\end{figure}

Figure~\ref{fig:massVsMult} shows the distribution of \mdv\ vs. \ntdv\ for vertices in the selected data
events, including vertices
that are rejected by the selection on \mdv\ and \ntdv. The signal distribution for the MH
sample is also indicated. The signal region, corresponding to a minimum number of five tracks in a vertex  and a minimum vertex mass of 10~GeV, is also
shown.
No events
are observed in the signal region.

\begin{figure}[!h]
\begin{center}
  \includegraphics[width=9cm]{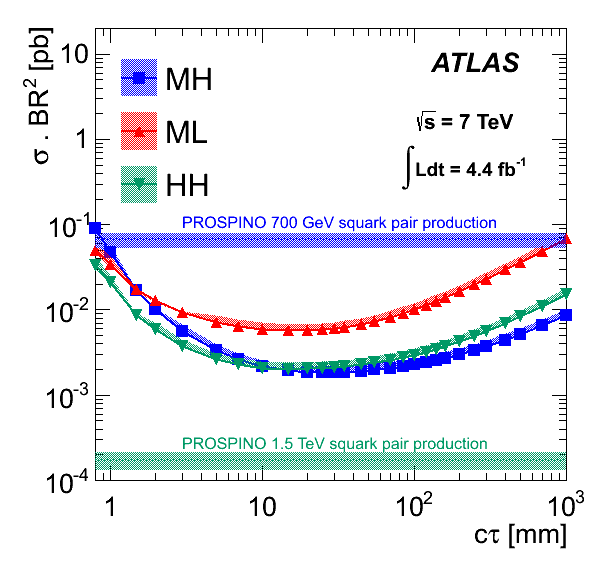}
  \caption{\label{fig:limitsn}
Upper limits at 95\% confidence level on $\sigma \cdot \mathrm{BR}^2$ vs. the neutralino lifetime for different combinations of
squark and neutralino masses, based on the observation of zero events
satisfying all criteria in a \lumi\ data sample. The shaded areas
around these curves represent the $\pm 1 \sigma$ uncertainty bands on
the expected limits. The horizontal lines
show the cross sections calculated from {\sc PROSPINO} for squark masses of
700~\gev~and 1500~\gev. The shaded regions around these lines
represent the uncertainties on the cross sections obtained from the
procedure described in the text.}
\end{center}
\end{figure}

Given the lack of candidate events in data,  upper limits are evaluated on the supersymmetry production
cross section ($\sigma$) times the square of the branching ratio
(BR$^2$) for produced squarks to decay via long-lived neutralinos to muons and quarks. The limits
are presented  for different assumed values of squark and neutralino mass and $c\tau$, where $\tau$
is the neutralino lifetime.  Based on the efficiency, background estimate, and luminosity of \lumi (with
the uncertainties on each of these treated as nuisance parameters), the $CL_s$ method~\cite{CLs} is used to set limits on the plane of
$\sigma \cdot \mathrm{BR}^2$  versus $c\tau$. The resulting limits are shown in Fig.~\ref{fig:limitsn}, with
the {\sc PROSPINO}-calculated cross sections for squark masses of 700 ~\gev~and 1500~\gev. The {\sc PROSPINO} predictions are shown
as a band which represents the variation in predicted cross section when using two different sets
of PDFs (CTEQ6.6 and MSTW2008NLO~\cite{Martin:2009iq}) and varying the factorisation and renormalisation
scales each up and down by a factor of two.  Since essentially no
background is expected and no events are observed,
the expected and observed limits are indistinguishable. Based on the observation of no
signal events in a data sample of $\lumi$, a 95\% confidence-level
upper limit of 0.68~fb is set on $\sigma \cdot \mathrm{BR}^2$ multiplied
by the detector acceptance and the reconstruction efficiency for any signal vertex.

\section{Summary and conclusions}
\label{sec:conclusions}

An improved search is presented for new, heavy particles that decay at
radial distances
between 4~\mm\ and 180~\mm\ from the $pp$
interaction point, in association with a high-transverse-momentum muon. This search is based on data collected by the ATLAS detector at the LHC at a center-of-mass energy of
7~TeV. Fewer than 0.06 background events are expected in the data sample
of \lumi, and no events are observed. Limits are derived on the
product of di-squark production cross section and decay chain
branching fraction squared, in a
SUGRA scenario where the lightest neutralino produced in the primary-squark decay undergoes an \RP~violating decay into a muon
and two quarks. The limits are
reported as a function of the neutralino lifetime and for a range of neutralino masses and velocities, which are
the factors with greatest impact on the reconstruction efficiency.
Limits for a variety of other models can thus be approximated from these results, based on
the neutralino mass and velocity distribution in a given model.
Comparing to the results of the previous analysis reported
in~\cite{Aad:2011zb}, for lifetimes $c\tau < 100\mathrm{mm}$, the limits presented here are somewhat less
stringent than could be expected from the factor $\approx 130$
increase in integrated luminosity, as a result of tighter trigger and
offline muon selection requirements. For longer lifetimes, these
factors are more than outweighed by the increase in efficiency
afforded by re-tracking, so the limits are somewhat better than would
be expected by simple scaling according to the integrated luminosity of the data samples.

We thank CERN for the very successful operation of the LHC, as well as the
support staff from our institutions without whom ATLAS could not be
operated efficiently.

We acknowledge the support of ANPCyT, Argentina; YerPhI, Armenia; ARC,
Australia; BMWF, Austria; ANAS, Azerbaijan; SSTC, Belarus; CNPq and FAPESP,
Brazil; NSERC, NRC and CFI, Canada; CERN; CONICYT, Chile; CAS, MOST and NSFC,
China; COLCIENCIAS, Colombia; MSMT CR, MPO CR and VSC CR, Czech Republic;
DNRF, DNSRC and Lundbeck Foundation, Denmark; EPLANET and ERC, European Union;
IN2P3-CNRS, CEA-DSM/IRFU, France; GNAS, Georgia; BMBF, DFG, HGF, MPG and AvH
Foundation, Germany; GSRT, Greece; ISF, MINERVA, GIF, DIP and Benoziyo Center,
Israel; INFN, Italy; MEXT and JSPS, Japan; CNRST, Morocco; FOM and NWO,
Netherlands; RCN, Norway; MNiSW, Poland; GRICES and FCT, Portugal; MERYS
(MECTS), Romania; MES of Russia and ROSATOM, Russian Federation; JINR; MSTD,
Serbia; MSSR, Slovakia; ARRS and MVZT, Slovenia; DST/NRF, South Africa;
MICINN, Spain; SRC and Wallenberg Foundation, Sweden; SER, SNSF and Cantons of
Bern and Geneva, Switzerland; NSC, Taiwan; TAEK, Turkey; STFC, the Royal
Society and Leverhulme Trust, United Kingdom; DOE and NSF, United States of
America.

The crucial computing support from all WLCG partners is acknowledged
gratefully, in particular from CERN and the ATLAS Tier-1 facilities at
TRIUMF (Canada), NDGF (Denmark, Norway, Sweden), CC-IN2P3 (France),
KIT/GridKA (Germany), INFN-CNAF (Italy), NL-T1 (Netherlands), PIC (Spain),
ASGC (Taiwan), RAL (UK) and BNL (USA) and in the Tier-2 facilities
worldwide.
\bibliographystyle{atlasnote}
\bibliography{DV-conf}

\clearpage

\begin{flushleft}
{\Large The ATLAS Collaboration}

\bigskip

G.~Aad$^{\rm 48}$,
T.~Abajyan$^{\rm 21}$,
B.~Abbott$^{\rm 111}$,
J.~Abdallah$^{\rm 12}$,
S.~Abdel~Khalek$^{\rm 115}$,
A.A.~Abdelalim$^{\rm 49}$,
O.~Abdinov$^{\rm 11}$,
R.~Aben$^{\rm 105}$,
B.~Abi$^{\rm 112}$,
M.~Abolins$^{\rm 88}$
% ...
\bigskip
\\
% ...
$^{11}$ Institute of Physics, Azerbaijan Academy of Sciences, Baku, Azerbaijan\\
$^{12}$ Institut de F{\'\i}sica d'Altes Energies and Departament de F{\'\i}sica de la Universitat Aut{\`o}noma de Barcelona and ICREA, Barcelona, Spain\\
% ...
$^{21}$ Physikalisches Institut, University of Bonn, Bonn, Germany\\
% ...
$^{48}$ Fakult{\"a}t f{\"u}r Mathematik und Physik, Albert-Ludwigs-Universit{\"a}t, Freiburg, Germany\\
$^{49}$ Section de Physique, Universit{\'e} de Gen{\`e}ve, Geneva, Switzerland\\
% ...
$^{88}$ Department of Physics and Astronomy, Michigan State University, East Lansing MI, United States of America\\
% ...
$^{105}$ Nikhef National Institute for Subatomic Physics and University of Amsterdam, Amsterdam, Netherlands\\
% ...
$^{111}$ Homer L. Dodge Department of Physics and Astronomy, University of Oklahoma, Norman OK, United States of America\\
$^{112}$ Department of Physics, Oklahoma State University, Stillwater OK, United States of America\\
% ...
$^{115}$ LAL, Universit{\'e} Paris-Sud and CNRS/IN2P3, Orsay, France\\
% ...
\end{flushleft}


\end{document}